\documentclass[a4paper,fleqn]{cas-sc}

\usepackage{tikz}
\usepackage{amsmath}   
\usepackage{placeins}
\usepackage{float}
\usepackage[numbers]{natbib}
\usepackage{subcaption} 

\usepackage{xurl}      
\usepackage{hyperref}

\usepackage{longtable} 
\usepackage{tabularx} 

\newcolumntype{Y}{>{\raggedright\arraybackslash}X}

\def\tsc#1{\csdef{#1}{\textsc{\lowercase{#1}}\xspace}}
\tsc{WGM}
\tsc{QE}

\begin{document}
\let\WriteBookmarks\relax
\def\floatpagepagefraction{1}
\def\textpagefraction{.001}

\shorttitle{}    

\shortauthors{}  


\title [mode = title]{SMoRFFI: A Large-Scale Same-Model 2.4 GHz Wi-Fi Dataset and Reproducible Framework for RF Fingerprinting}

\author[1]{Zewei Guo}


\ead{zwguo1996@gmail.com}

\author[2]{Zhen Jia}
\ead{jiazhen0628@outlook.com}

\author[3]{JinXiao Zhu}
\ead{jyuu@mail.dendai.ac.jp}

\author[2]{Wenhao Huang}
\ead{gerhua@sfc.keio.ac.jp}

\author[4,2]{Yin Chen}
\cormark[1]
\ead{ychen@reitaku-u.ac.jp}

\affiliation[1]{organization={Future University Hakodate},
            city={Hakodate},
            postcode={041-8655}, 
            state={Hokkaido},
            country={Japan}}

\affiliation[2]{organization={Keio University},
            city={Fujisawa},
            postcode={108-8345}, 
            state={Kanagawa},
            country={Japan}}
\affiliation[3]{organization={Tokyo Denki University},
            city={Tokyo},
            postcode={120-8551}, 
            state={Tokyo},
            country={Japan}}
\affiliation[4]{organization={Reitaku University},
            city={Kashiwa},
            postcode={277-8686}, 
            state={Chiba},
            country={Japan}}

\cortext[1]{Corresponding author}



\begin{abstract}
Radio frequency (RF) fingerprinting exploits hardware imperfections for device identification, but distinguishing between same-model devices remains challenging due to their minimal hardware variations. Existing datasets for RF fingerprinting are constrained by small device scales and heterogeneous models, which hinder robust training and fair evaluation of machine learning methods. To address this gap, we introduce a large-scale dataset of same-model devices along with an open-source experimental framework. The dataset is built using 123 same-model commercial IEEE 802.11g devices, which contain 35.42 million raw I/Q samples from the preambles and corresponding 1.85 million RF features. The accompanying framework further provides a fully reproducible pipeline from data collection to performance evaluation. Within this framework, a Random Forest–based algorithm is implemented as a baseline to achieve 89.06\% identification accuracy on this dataset. 

\end{abstract}

\maketitle

\begin{tabular}{p{0.24\textwidth}p{0.71\textwidth}}  
\\
\multicolumn{2}{l}{\hspace{-2mm}\vspace{2mm}\textbf{Specifications table}} \\
\hline
Subject & Wireless communication \\
\hline
Specific subject area & Radio fingerprinting techniques based on machine learning algorithms. \\
\hline
Type of data & CSV files. \\
\hline
{How data were acquired} &  This dataset was acquired using $1$ USRP B$210$ receiver operated via GNU Radio, along with $123$ M5Stack Core2 transmitters, $1$ Huawei WS7100 V2 Wi-Fi access point, and $1$ Minisforum NPB7 computer. \\
\hline
{Data format} & I/Q samples of preamble and corresponding radio frequency (RF) features. Specifically, each recorded frame mainly contains the following elements: (1) MAC address of the device, (2) Raw samples of preamble,
(3) I/Q samples of long preamble, (4) Coarse Carrier frequency offset (CFO), (5) Fine CFO, (6) Final CFO, (7) Phase error, (8) Magnitude error, (9) I/Q gain imbalance, (10) Fractal dimension. \\
\hline
{Parameters for data collection} & $1,000$ signal frames were collected from each transmitter. All transmitters and the USRP B$210$ receiver were set to work under the IEEE $802.11$g standard with $20$ MHz bandwidth. The receiver used the IEEE 802.11 a/g/p transceiver model~\cite{Bloessl2013}, and the sampling rate was set as $20$ MS/s.\\
\hline
{Description of data collection} &   To extract raw I/Q samples of the preamble, customized modifications were conducted on the IEEE 802.11 a/g/p transceiver model~\cite{Bloessl2013}. Specifically, the "WiFi Sync Short", "WiFi Sync Long", and "WiFi Frame Equalizer" blocks were rewritten, and the modified "RFtap Encapsulation" block~\cite{RFtap2016} was integrated. These customized models were then used to receive Wi-Fi signals and record the raw preamble samples. \\

\hline
\multicolumn{2}{r}{\textit{(continued on next page)}} \\
\end{tabular}

\begin{table}[t]
\renewcommand{\arraystretch}{1.3}
\centering
{\normalsize\normalfont  
\begin{tabular}{p{0.24\textwidth}p{0.71\textwidth}}
\multicolumn{2}{l}{\textit{(continued)}} \\
\hline
Data source location & 
\begin{tabular}[t]{@{}l@{}}
Institution: Reitaku University \\
City: Kashiwa, Chiba \\
Country: Japan
\end{tabular} \\
\hline
Data accessibility & 
\begin{minipage}[t]{\linewidth}\raggedright
Repository name: RFFI-IQ\_only-wifi-802.11g-2.4G-123-m5stack\\
Direct URL to data: \url{https://www.kaggle.com/datasets/yinchen1986/rffi-123-m5stack-iq-wifi-802-11g-2-4g}\\
Repository name: RFFI-kf\_feature-IQ-wifi-802.11g-2.4G-123-m5stack\\
Direct URL to data: \url{https://www.kaggle.com/datasets/yinchen1986/rffi-kf-feature-iq-wifi-802-11g-2-4g-123-m5stack}
\end{minipage}\\
\hline
Related code accessibility & 
\begin{minipage}[t]{\linewidth}\raggedright
Repository name: Dockerized-wifi-iq-preamble-capture\\
Direct URL to code: \url{https://github.com/aiot-lab-yin/Dockerized-wifi-iq-preamble-capture}\\[3pt]
Repository name: RFF\_Data\_Calculate \\
Direct URL to code: \url{https://www.kaggle.com/code/zeweiguo/rff-data-calculate}
\end{minipage} \\
\hline
{Related research article(s)} & 
\begin{tabular}[t]{@{}l@{}}
Author's name: Zhen Jia, Zewei Guo, Wenhao Huang, Yin Chen, Jinxiao Zhu \\ \hspace{22.5mm}and Xiaohong Jiang  \\
Title: An Experimental Study on Radio Frequency Fingerprinting-based Authentication \\\hspace{8.5mm}in IEEE 802.11g  \\
Journal: IEICE Tech. Rep., vol. 124, no. 122, BioX2024-70, pp. 344-349, Jul. 2024 \\
\end{tabular} \\
\hline
\end{tabular}
} 
\label{tab:metadata-continued}
\end{table}

\vspace{5pt}
\emph{Notation}: 
Lower-case and lower-case bold-face letters represent scalars and vectors (e.g., $a$ and $\mathbf{a}$), respectively. Let $||\mathbf{a}||$ and ${\mathbf a}^\dagger$ denote the Frobenius norm and conjugate transpose of vector ${\mathbf a}$, respectively.
Denote $\Re\{\mathscr{c}\}$, $\Im\{\mathscr{c}\}$, $|\mathscr{c}|$, and $\angle(\mathscr{c})$  as the real part, the imaginary part, the absolute value, and the phase of the complex variable $\mathscr{c}$. 

\section{Value of the Data}

\subsection{Why are these data useful?}
RF fingerprinting has recently attracted increasing attention from both academia and industry ~\cite{Zeng2022,Zeng2023}. Although several public RF feature datasets are available~\cite{Sankhe2019,Uzundurukan2020,AlShawabka2020,Hanna2022}, these datasets have several limitations: (1) they typically contain a limited number of devices and small-scale samples, which are insufficient for training effective machine learning models; 
(2) the devices in these datasets are of different models, which makes it difficult to verify the effectiveness of identification methods on a large number of same-model devices;
(3) most existing datasets only provide raw I/Q samples and require additional processing to extract relevant RF features.
To address these limitations, we introduce a dataset that integrates large scale, high device homogeneity, and comprehensive feature coverage. The dataset contains 35.42 million raw I/Q samples and 1.85 million feature records collected from 123 devices of the same model. Since same-model devices produce much smaller RF differences, such a large homogeneous group represents a worst-case scenario for RF fingerprinting and therefore provides a more stringent and informative benchmark for future studies.

\subsection{Who can benefit from these data?} 
Developers who work on RF fingerprinting algorithms will find this dataset particularly valuable. In addition, data scientists who focus on the exploration and analysis of RF features can use this dataset for their research purposes.

\subsection {How can these data be used for further insights and development of experiments?} 

\begin{itemize}
    \item The dataset serves as a large-scale benchmark that allows RF feature-based algorithm developers to train and evaluate machine learning and deep learning models.
    \item It provides extensive data for researchers to verify the effectiveness of newly proposed features.
    \item Preliminary analyzes of the extracted RF features are also included, which help researchers compare their new features with those contained in this dataset.
    \item A complete experimental framework is provided to support reproducible research, which covers the entire process from data collection to performance evaluation. It allows researchers to develop and extend their own experiments based on this dataset.
\end{itemize}

\subsection{What is the additional value of these data?} 
In addition to the I/Q samples and the corresponding RF features, this paper proposes a baseline physical layer identification algorithm based on the Random Forest model and provides related performance evaluations.
This baseline allows developers to compare the performance of their RF feature-based algorithms with a standardized reference.

\section{Data}
\label{dataset}
\FloatBarrier
\begin{figure}[t]
    \centering
    \includegraphics[width=0.85\textwidth]{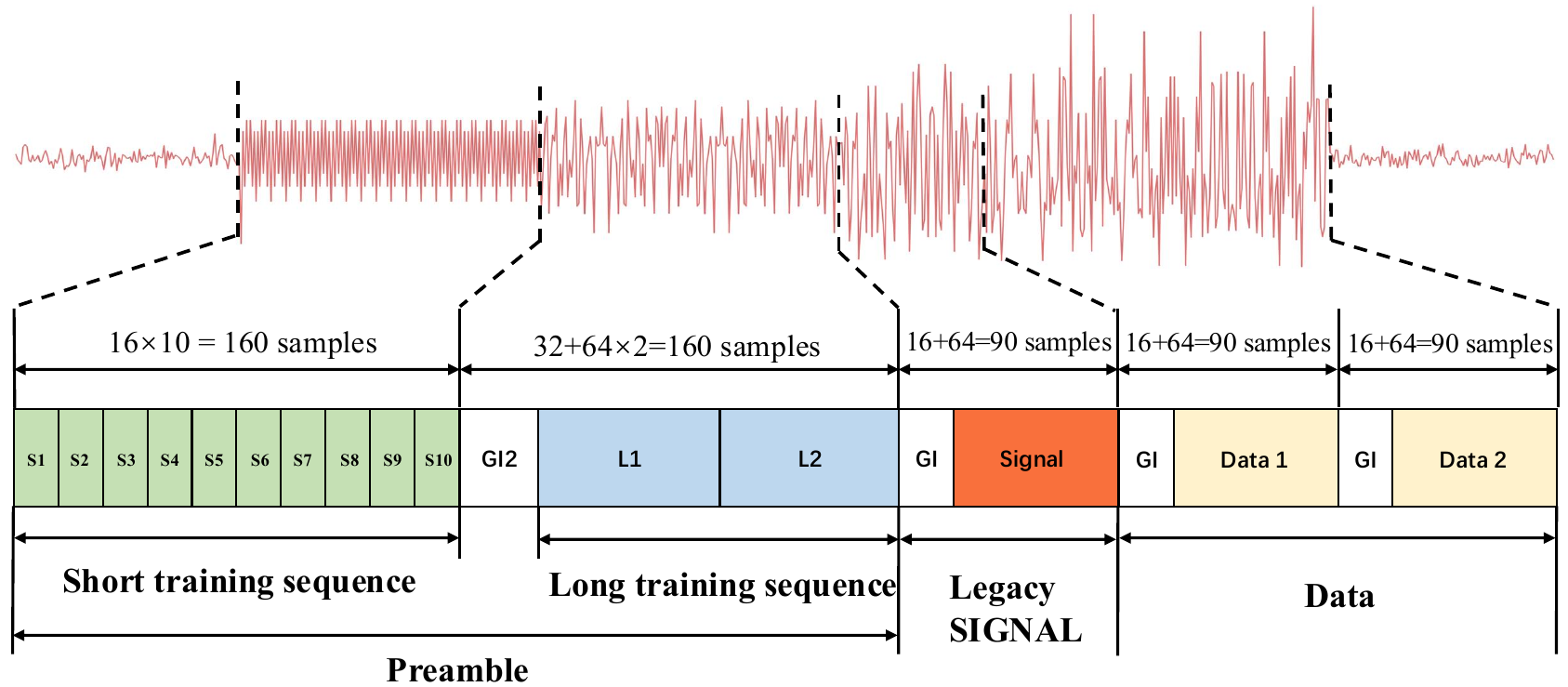}
    \caption{Structure of the IEEE 802.11g frame \cite{Committee2009}.}
    \label{FIG:preamble}
\end{figure}

This paper provides two types of datasets. Each dataset contains 123 CSV files, and each file includes 1,000 records collected from a single IoT device.
The first dataset (named "RFFI-IQ\_only-wifi-802.11g-2.4G-123-m5stack") includes only the MAC address and the preamble samples from each device.
The second dataset (named "RFFI-kf\_feature-IQ-wifi-802.11g-2.4G-123-m5stack") provides multi-dimensional RF features, where each record consists of ten categories of information described as follows:

 (1) {\bf MAC address (mac\_address)} serves as the unique label for each IoT device and is used to distinguish among different devices.
 
 (2) {\bf Raw samples of the preamble (preamble)} refer to the preamble data that have not undergone any processing steps.
 The structure of the preamble is defined in the IEEE 802.11g standard, and its location within the signal frame is shown in Figure~\ref{FIG:preamble}.
 As shown in Figure~\ref{FIG:preamble}, the preamble consists of a short training sequence (STS), a guard interval (GI2), and a long training sequence (LTS).
 Specifically, the STS consists of 10 identical training sequences, and each sequence has a length of $16$ samples. These sequences are used by the receiver for packet detection and coarse CFO correction. 
 LTS includes two identical training sequences (i.e., L1 and L2), where each sequence consists of 64 samples. These sequences are used for fine carrier frequency offset (CFO) correction.
 GI2 is a cyclic prefix sequence consisting of 32 samples, which is used to prevent interference between the STS and LTS.
 To prevent STS truncation caused by erroneous packet detection, the first two training sequences of STS are dropped in this dataset. Consequently, the resulting preamble sequence ${\mathbf{x}}$ can be expressed as
\begin{equation}
    \mathbf{x} = [
    \underbrace{x_1, \ldots, x_{128}}_{\text{STS}},
     \ldots, 
    \underbrace{x_{161}, \ldots, x_{288}}_{\text{LTS}}
    ].
\end{equation}


 (3) {\bf Coarse CFO (short\_freq)} is the CFO estimated using STS, which can be expressed as \cite{Sourour2004}
\begin{equation}
{\alpha}_{S}=\frac{1}{16} \angle\left(\sum_{i=1}^{112} \mathbf{x}[i]^\dagger \mathbf{x}[i+16]\right).
\end{equation}
After obtaining ${\alpha}_{S}$, the I/Q samples of LTS are corrected as
\begin{equation}
    \hat{\mathbf{x}}[i] = \mathbf{x}[160+i]e^{-\jmath(i-1){\alpha}_{S}}, i=1,2,\cdots, 128,
\end{equation}
where $\jmath=\sqrt{-1}$.

 (4) {\bf Fine CFO (long\_freq)} is the CFO estimated from LTS after coarse CFO correction, which can be calculated as \cite{Sourour2004}
\begin{equation}
{\alpha}_{L}=\frac{1}{64} \angle\left(\sum_{i=1}^{64} \hat{\mathbf{x}}[i]^\dagger \hat{\mathbf{x}}[i+64]\right).
\end{equation}
Then, the fine CFO correction can be expressed as
\begin{equation}
    \check{\mathbf{x}}[i] = \hat{\mathbf{x}}[i]e^{-\jmath(i-1){\alpha}_{L}}, i=1,2,\cdots128.
\end{equation}

(5) {\bf Final CFO (CFO)} denotes the total CFO of the signal frame. It is applied to compensate for the CFO of the remaining signal after the LTS. The final CFO can be expressed as \cite{Sourour2004}
\vspace{-10pt}
\begin{equation}
    {\alpha}_{F}={\alpha}_{S}+{\alpha}_{L}.
\end{equation}

(6) {\bf Measured LTS (iq\_preamble)} denotes the LTS samples after frequency offset correction, Fourier transform, and sub-carrier equalization, which can be given by \cite{openOFDM}
\begin{equation}
    \mathbf{l}[i] = \frac{\check{\mathbf{x}}_{\text{FFT}}[i]}{\mathbf{h}[i\bmod 64]},i=1,2,\cdots128.
\end{equation}
Here, $\check{\mathbf{x}}_{\text{FFT}}[i]$ denotes the LTS samples after the $64$ point fast Fourier transform (FFT). $\mathbf{h}[k]$ denotes the channel gain, which is expressed as
\begin{equation}
    \mathbf{h}[j] = \frac{1}{2}\left(\check{\mathbf{x}}[j]+\check{\mathbf{x}}[64+j]\right) \times \mathbf{l}_{\text{ideal}}[j],j=1,2,\cdots64,
\end{equation}
where $\mathbf{l}_{\text{ideal}}$ is the ideal LTS symbol defined in [\citenum{Committee2009}, Table L-5] and can be expressed as
\begin{equation}
\begin{aligned}
\mathbf{l}_{\text{ideal}} = [&0,0,0,0,0,0,1,1,-1,-1,1,1,-1,1,-1,1,1,\\
&1,1,1,1,-1,-1,1,1,-1,1,-1,1,1,1,1,0,1, \\
&-1,-1,1,1,-1,1,-1,1,-1,-1,-1,-1,-1,\\
&1,1,-1,-1,1,-1,1,-1,1,1,1,1,0,0,0,0,0].
\end{aligned}
\end{equation}
Finally, the two identical training sequences contained in the long preamble, denoted as L1 and L2, are represented by $\mathbf{l}_1$ and $\mathbf{l}_2$, respectively.
 
 (7) {\bf Phase error vector (phase\_error\_v1 and phase\_} {\bf error\_v2)} denotes the angle between the measured and ideal phasors, which can be calculated as \cite{Brik2008}
\begin{equation}
\mathbf{p}_{k}[j] = \angle\mathbf{l}_{k}[j] - \angle\mathbf{l}_{\text{ideal}}[j],k\in\{1,2\},j=1,2,\cdots64.
\end{equation}
The mean and variance of the phase error vector are further computed and recorded in the dataset as {phase\_error\_mean} and {phase\_error\_var}, respectively.
 
 (8) {\bf Magnitude error vector (mag\_error\_v1 and mag\_} {\bf error\_v2)} is the difference in the magnitudes of the ideal and measured phasors, which can be expressed as~\cite{Brik2008}
  \begin{equation}
     \mathbf{m}_{k}[j] = |\mathbf{l}_k[j]|-|\mathbf{l}_{\text{ideal}}[j]|,k\in\{1,2\},j=1,2,\cdots64.
 \end{equation}
The mean and variance of each magnitude error vector are also computed and recorded in the dataset as {\tt mag\_error\_mean} and {\tt mag\_error\_var}, respectively.
 
(9) {\bf I/Q gain imbalance (iqi\_1 and iqi\_2)} refers to the gain imbalance in the in-phase (I) and quadrature (Q) components of a signal, which can be expressed as \cite{Arslan2006}
\begin{equation}
    IQI_{k}= \frac{\|\Re\{ \mathbf{l}_k\}\|^2}{\|\Im\{ \mathbf{l}_k\}\|^2},k\in\{1,2\}.
\end{equation}

\FloatBarrier
\begin{figure}[thb]
    \centering
    \includegraphics[width=1\textwidth]{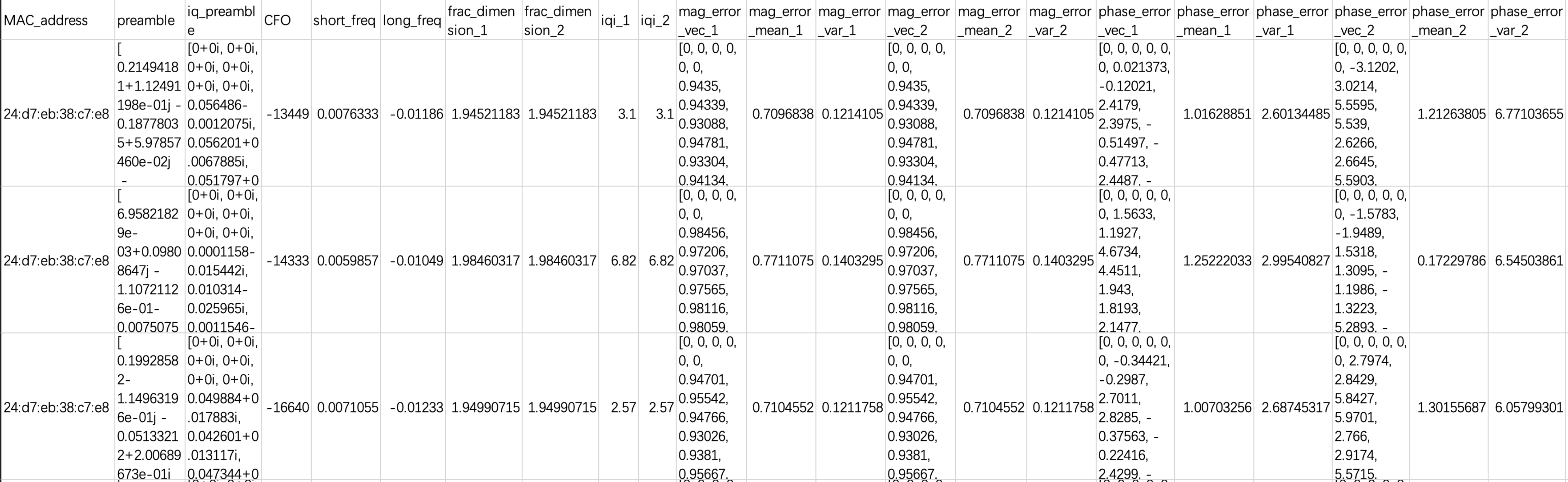}
    \caption{Structure of the file 24:d7:eb:38:c7:e8\_pre.csv.}
    \label{FIG:CSV_table_structure}
\end{figure}

(10) {\bf Fractal dimension (frac\_dimension\_1 and frac\_} {\bf dimension\_2)} quantifies the complexity and self-similarity of the I/Q signal distribution. It can serve as a statistical measure of the geometric structure in the sampled data. 
Following the procedure in \cite{Li2024}, the received LTS samples are processed into the same type of symbol to facilitate the calculation of the fractal dimension. 
Specifically, for each sample of the LTS $\mathbf{l}[n]$, it will be subtracted by its corresponding ideal value $\mathbf{l}_{\text{ideal}}[n]$.
The processed LTS samples are denoted by $\hat{\mathbf{l}}$, and the samples of the two identical training sequences are denoted by $\hat{\mathbf{l}}_1$ and $\hat{\mathbf{l}}_2$, respectively. Consequently, the corresponding fractal dimension $\mathcal{D}$ is estimated as 
\begin{equation}
\mathcal{D}_{k}=\frac{1}{2} \times\left(3-\frac{\ln \sum_{n=1}^{\lfloor\mathrm{64} / \tau\rfloor} d_{\tau, n,k}-\ln \sum_{n=1}^{\lfloor\mathrm{64} / \tau\rfloor \cdot \tau} d_{1, n,k}}{\ln \tau}\right),
\end{equation}
where
\begin{align}
    d_{\tau, n,k}&=\left\|\hat{\mathbf l}_k[n \cdot \tau]-\hat{\mathbf l}_k[(n-1) \cdot \tau]\right\|,\\
    d_{1, n,k}&=\left\|\hat{\mathbf l}_k[n]-\hat{\mathbf l}_k[n-1]\right\|,k\in\{1,2\}.
\end{align}
Here, $\tau$ is a scale parameter, $d_{\tau, n, k}$ denotes the Euclidean distance between the $(n-1)\tau$-th and $n\cdot\tau$-th symbols, and $d_{1, n, k}$ denotes the Euclidean distance between the $(n-1)$-th and $n$-th symbols. Following \cite{Li2024}, the scale parameter is set to $\tau = 3$ in this dataset.

Based on the above definitions of RF features, the preamble and related RF features of each IoT device can be calculated and organized into CSV files. The structure of an example file is shown in Figure \ref{FIG:CSV_table_structure}.



\section{Experimental Design, Materials and Methods}
\label{sec:framework}
\subsection{Dataset construction workflow}
\begin{figure}[tbp]
    \centering
    \includegraphics[width=0.5\linewidth]{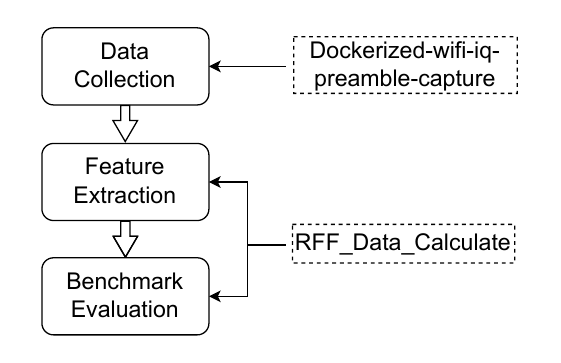}
    \caption{The workflow for constructing the dataset.}
    \label{framework}
\end{figure}

This section presents the overall workflow for constructing the dataset.
As shown in Figure \ref{framework}, the framework is composed of three modules: (1) data collection, (2) feature extraction, and (3) benchmark evaluation.
The data collection module collects I/Q samples of preambles from 123 same-model devices.
Then, the feature extraction module calculates the RF features from the collected I/Q samples.
By utilizing these features, the benchmark evaluation module provides reproducible baseline results.

\subsection{Data collection}
\label{datacollection}

\begin{figure}[tbp]
    \centering
    \includegraphics[width=0.475\textwidth]{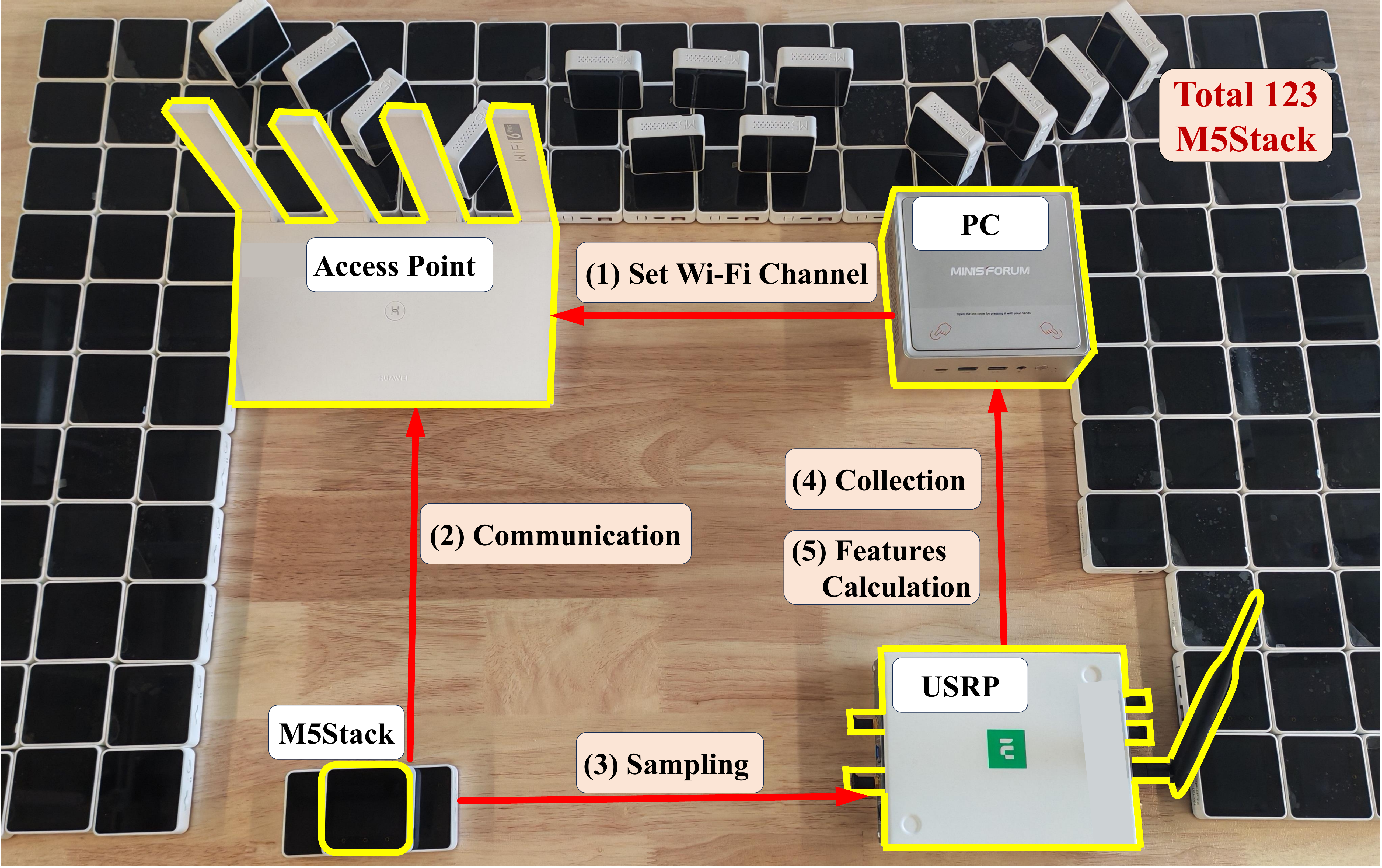}
    \caption{Experimental platform.}
    \label{FIG:system_model}
\end{figure}

As shown in Fig.~\ref{FIG:system_model}, the collection platform consists of a USRP~B210 software-defined radio acting as a receiver, 123 M5Stack Core2 devices serving as transmitters, a Huawei WS7100~V2 Wi-Fi access point (AP), and a Minisforum NPB7 computer. AP is configured according to the IEEE 802.11g standard on Channel~6 (20~MHz bandwidth). The computer runs the GNU Radio 802.11 a/g/p transceiver model~\cite{Bloessl2013}, which controls the USRP and manages baseband signal processing. All experiments are conducted in an indoor office with a fixed 50~cm separation between each transmitter and the receiver.

The data collection procedure can be summarized as follows.  
First, the AP is configured to operate on the target Wi-Fi channel, ensuring that all transmissions from the transmitters are routed through this channel. A Python control script then triggers a transmitter to repeatedly transmit packets toward the AP. During transmissions, the USRP continuously listens and samples the preambles of signals. The captured samples are then stored on the computer as a CSV file. This process is repeated until 1,000 preambles have been collected for each device.

\subsection{Feature extraction}
\label{Feature Extraction}

\FloatBarrier
\begin{figure}[htbp]
    \centering
    \includegraphics[width=0.6\textwidth]{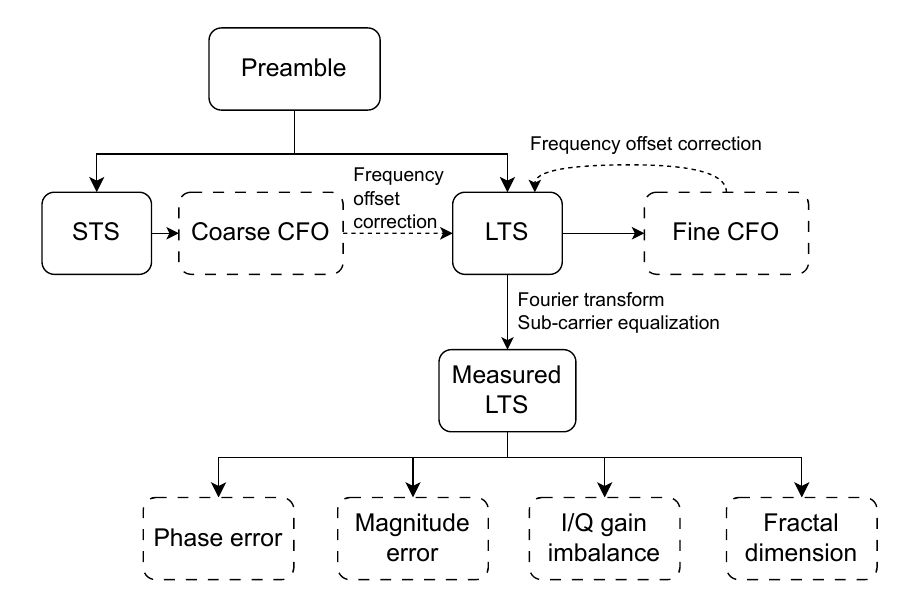}
    \caption{Workflow of the feature extraction model.}
    \label{FIG:feature_flow}
\end{figure}

Based on the raw I/Q preambles collected in the first module, this module extracts multi-dimensional feature representations from the sampled data. The workflow of the feature extraction model is illustrated in Figure~\ref{FIG:feature_flow}. Specifically, we first extract the coarse CFO and fine CFO from the STS and LTS, respectively. Then, frequency offset correction, Fourier transform, and subcarrier equalization are applied to the LTS to obtain the measured LTS. Finally, the phase error, magnitude error, I/Q gain imbalance, and fractal dimension are calculated based on the measured LTS.





\subsection{Benchmark evaluation}
\label{Benchmark Evaluation}
This module provides a unified environment for analyzing the extracted features and establishing a reproducible baseline for RF fingerprinting. 
It operates on the features produced by Section \ref{Feature Extraction} and supports both dataset enhancement and benchmark evaluation.

\textbf{Feature visualization}: To help users examine the characteristics of the extracted features, the framework includes t-SNE–based embedding tools and correlation-analysis utilities.  These visualizations offer an intuitive view of feature separability of transmitters.

\textbf{Feature enhancement}: This tool provides an optional Kalman Filtering–based smoothing mechanism that can be applied to reduce noise in the computed features.  

\textbf{Benchmark algorithm}:  A Random Forest classifier is implemented as the default identification baseline.  This benchmark provides a standardized entry point for evaluating new RF fingerprinting methods on the dataset.

\subsection{Implementation resources}
To support full reproducibility of the proposed framework, we provide open-source implementations that correspond directly to each module described above.

\textbf{Data collection environment.}
A complete data collection module is fully encapsulated in a Docker container. This containerized setup can be deployed to reproduce the procedure in Section \ref{datacollection}. Repository: 
\url{https://github.com/aiot-lab-yin/Dockerized-wifi-iq-preamble-capture}

\textbf{Feature extraction and benchmark evaluation.}
Both modules are encapsulated in a Jupyter Notebook using the Python programming language, which provides online implementation from feature extraction (Section~\ref{Feature Extraction}) to benchmark evaluation (Section~\ref{Benchmark Evaluation}).
Repository: 
\url{https://www.kaggle.com/code/zeweiguo/rff-data-calculate}


\section{Experimental Results and Analysis}
\subsection{Feature enhancement via Kalman Filtering}
The extracted RF features exhibit temporal variations caused by hardware imperfections, measurement noise, and minor environmental fluctuations. These variations blur the device-dependent patterns and reduce inter-device separability. To obtain a more stable representation, we apply Kalman Filtering (KF) to smooth each feature dimension.

\begin{figure}[htbp]
    \centering
    \begin{subfigure}{1\linewidth}
        \centering
        \includegraphics[width=0.6\linewidth]{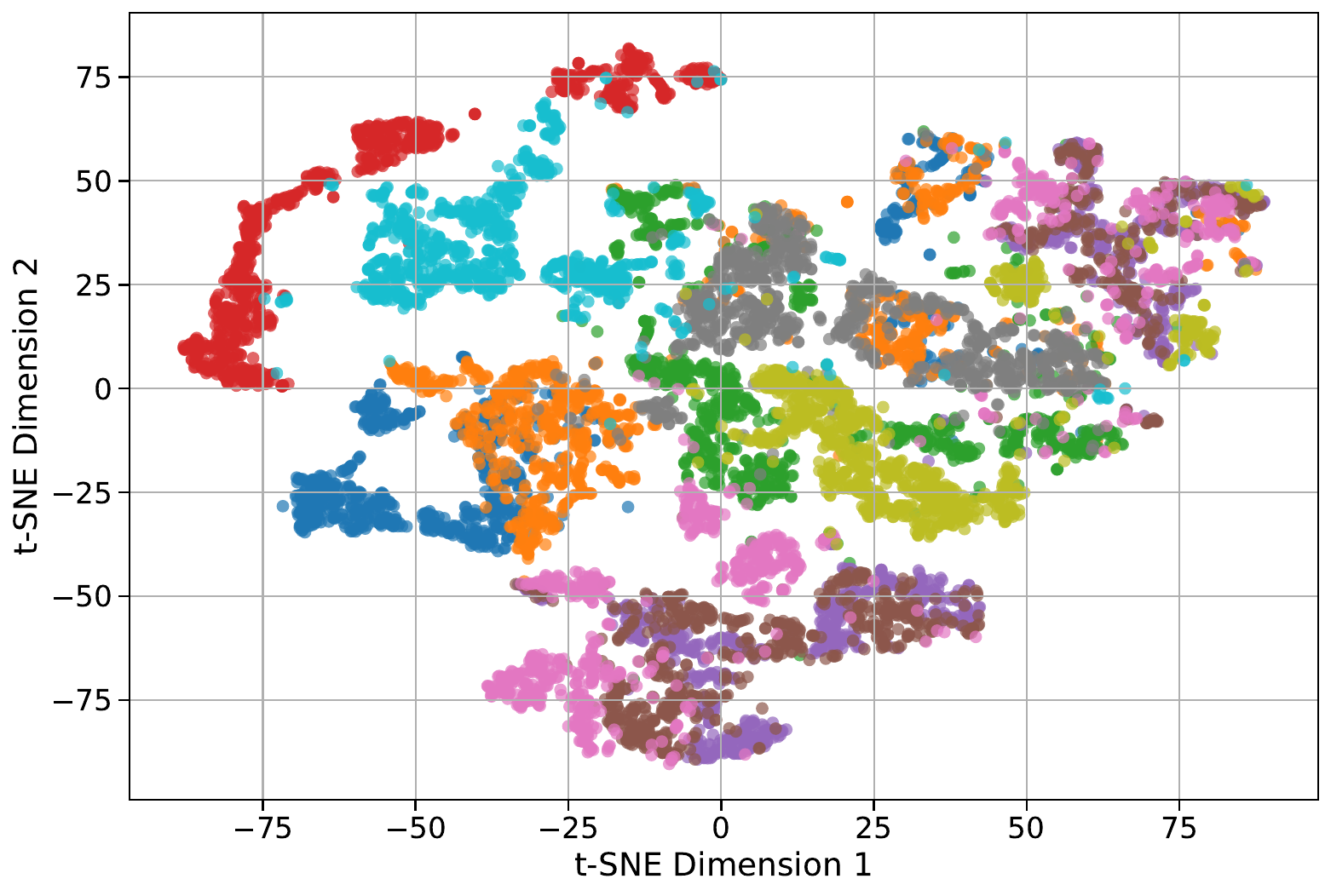}
        \caption{Before Kalman Filtering}
        \label{fig:before_KF}
    \end{subfigure}
    \hfill
    \begin{subfigure}{1\linewidth}
        \centering
        \includegraphics[width=0.6\linewidth]{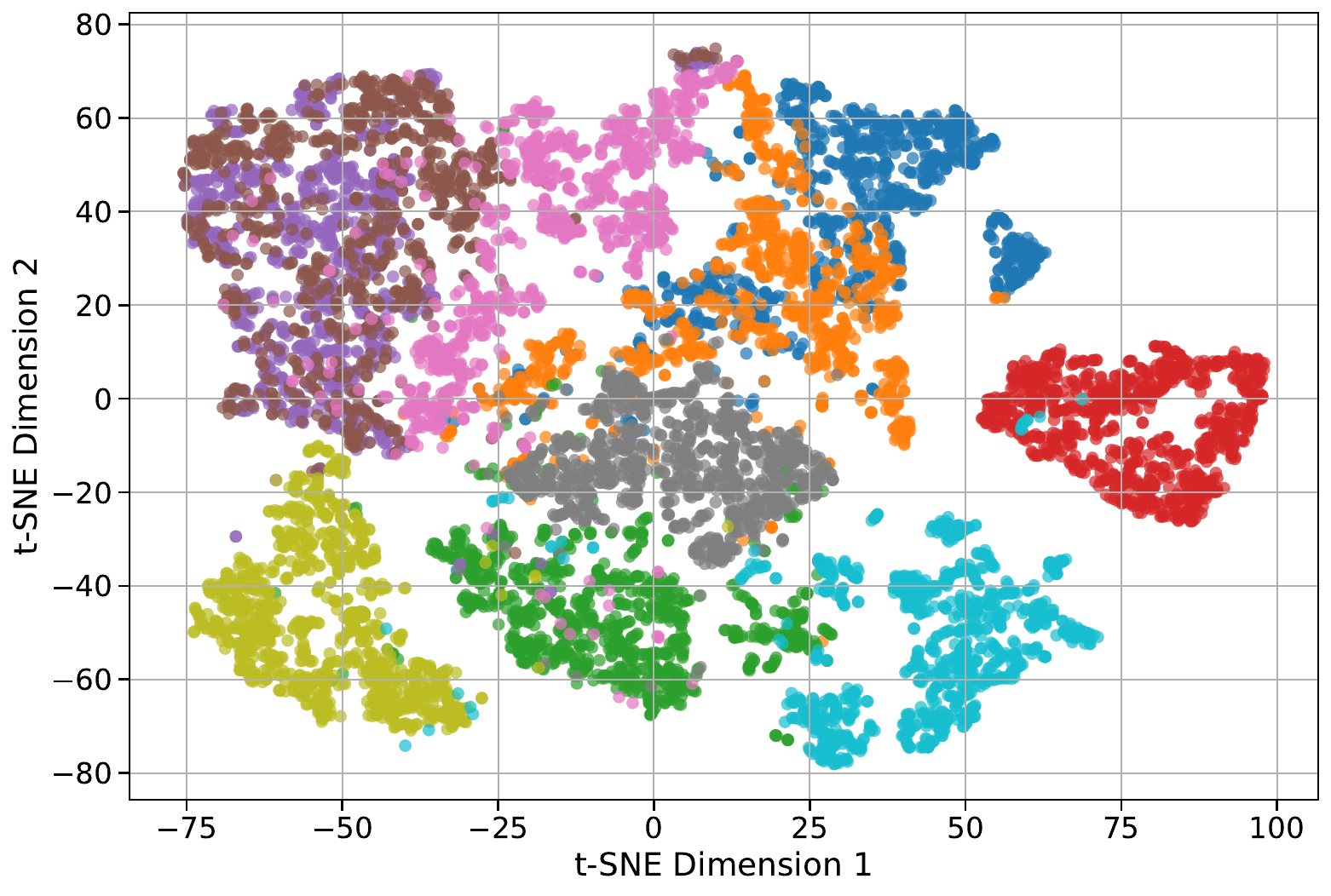}
        \caption{After Kalman Filtering}
        \label{fig:after_KF}
    \end{subfigure}
    \caption{RF fingerprint features for $10$ randomly selected devices, where each color represents a device.}
    \label{fig:KF}
\end{figure}

To demonstrate the effect of KF on this dataset, we visualize the high-dimensional feature vectors using t-distributed Stochastic Neighbor Embedding (t-SNE). For clarity, the visualization is generated from ten randomly selected devices, with each color representing one device. As shown in Figure~\ref{fig:before_KF}, the unfiltered features form clusters with noticeable overlap, indicating weakened device distinguishability. After applying KF, the clusters in Figure~\ref{fig:after_KF} become compact within each device and more clearly separated across devices, which confirms that KF improves the discriminability of the feature space.
\FloatBarrier

\subsection{Benchmark results}
\begin{table}[h!]
	\centering  
	\caption{Evaluation Results for Random Forest Model}  
	\label{table:ACC}  
        \renewcommand{\arraystretch}{1.2}
	\begin{tabular}{|@{\hskip 7pt}p{0.25\textwidth}@{\hskip 3pt}|>{\centering\arraybackslash}p{0.18\textwidth}@{\hskip 3pt}|}  
		\hline  
		\textbf{Dataset \& Setting} & \textbf{Accuracy (ACC)} \\  
		\hline
            5-Fold Average \textbf{without} KF & 82.18 \\
		\hline
            5-Fold Average \textbf{with} KF & 89.06 \\
        \hline
	\end{tabular}
\end{table}

In this subsection, we adopt the Random Forest model to achieve baseline identification performance on the proposed dataset. 
In particular, the Random Forest classifier is configured with $128$ decision trees, trained on $80\%$ of the data, and tested on the remaining $20\%$ to evaluate the performance. Based on the established baseline model, we further analyze the relative importance and inter-feature correlations among the extracted RF features.

As summarized in Table~\ref{table:ACC}, the 5-fold cross-validation results of the Random Forest model are reported for both cases with and without KF. Results show that applying KF increases the average accuracy from $82.18\%$ to $89.06\%$. This quantitative improvement is consistent with the enhancement in feature separation visualized in Figure~\ref{fig:after_KF}, which validates the effectiveness of KF in enhancing identification accuracy.

\begin{table}
	\centering  
	\caption{Feature Importance for Random Forest Model}  
	\label{table:Importance}  
        \renewcommand{\arraystretch}{1.2}
	\begin{tabular}{|@{\hskip 7pt}p{0.52\textwidth}@{\hskip 3pt}|>{\centering\arraybackslash}p{0.1\textwidth}@{\hskip 3pt}|}  
		\hline  
		\textbf{Feature} & \textbf{Importance} \\  
		\hline
            CFO & 0.2199 \\
		\hline
            Coarse CFO of STS (short\_freq) & 0.1760 \\
        \hline
            Fine CFO of LTS (long\_freq) & 0.1442 \\
        \hline
            Mean of Phase Error Vector of L1 (phase\_error\_mean\_1) & 0.0562 \\
        \hline
            Mean of Phase Error Vector of L2 (phase\_error\_mean\_2) & 0.0544 \\
        \hline
            Variance of Phase Error Vector of L2 (phase\_error\_var\_2) & 0.0418 \\
        \hline
            Variance of Phase Error Vector of L1 (phase\_error\_var\_1) & 0.0417 \\
        \hline
            I/Q Gain Imbalance of L1 (iqi\_1) & 0.0382 \\
        \hline
            Mean of Magnitude Error Vector of L2 (mag\_error\_mean\_2) & 0.0381 \\
        \hline
            Mean of Magnitude Error Vector of L1 (mag\_error\_mean\_1)  & 0.0375 \\
        \hline
            I/Q Gain Imbalance of L2 (iqi\_2) & 0.0375 \\
        \hline
            Variance of Magnitude Error Vector of L1 (mag\_error\_var\_1) & 0.0305 \\
        \hline
            Variance of Magnitude Error Vector of L2 (mag\_error\_var\_2) & 0.0305 \\
        \hline
            Fractal Dimension of L1 (frac\_dimension\_1) & 0.0268 \\
        \hline
            Fractal Dimension of L2 (frac\_dimension\_2) & 0.0265 \\
        \hline
	\end{tabular}
\end{table}

Following the accuracy evaluation, Table~\ref{table:Importance} presents the feature importance evaluation for device identification based on the Random Forest model. 
As shown in Table~\ref{table:Importance}, frequency-related features (i.e., CFO, short\_freq, and long\_freq) are the most discriminative, reflecting both their robustness and device-specific characteristics. Constellation-related features such as phase and magnitude-based features, together with I/Q gain imbalance, exhibit moderate importance. In contrast, the fractal dimension features rank lower but still contribute to device identification.

\begin{figure}
    \centering
    \includegraphics[width=0.9\linewidth]{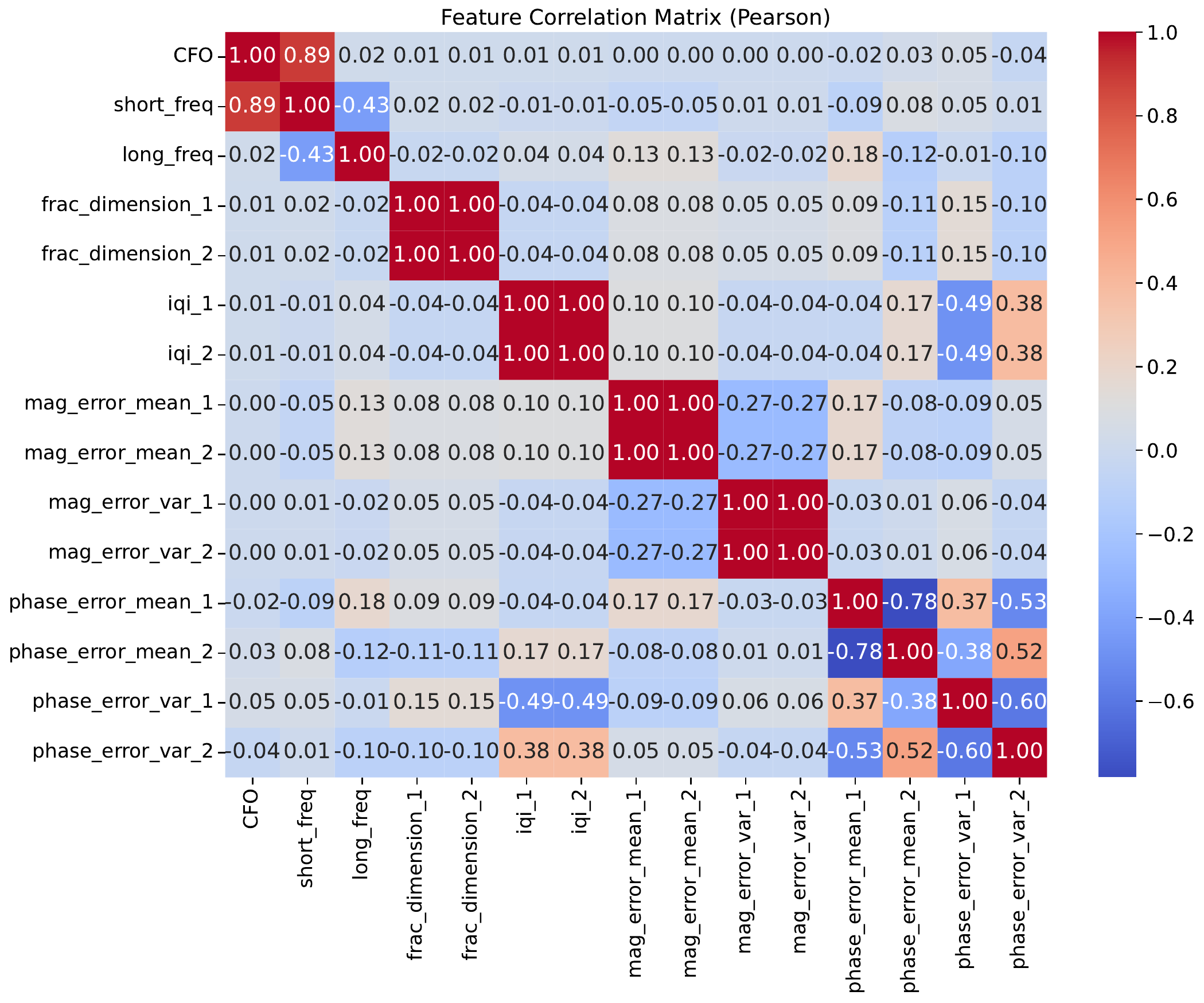}
    \caption{Correlation matrix of RF fingerprint features.}
    \label{fig:placeholder}
\end{figure}

To explore the relationships among the extracted features, the Pearson correlation matrix is illustrated in Figure~\ref{fig:placeholder}. We can observe a strong positive correlation ($r = 0.89$) between the CFO and the coarse CFO of the STS (short\_freq), which indicates that they capture similar characteristics. In contrast, the long\_freq feature shows negligible correlation with other frequency-related features, highlighting it provides additional independent information. Furthermore, the two fractal dimension features (frac\_dimension \_1 and frac\_dimension\_2) exhibit an extremely high correlation ($r = 1$), implying that they have consistent structural patterns and potential redundancy. Meanwhile, the constellation-related features (e.g., mag\_error\_mean, phase\_ error\_mean) show low correlations, which demonstrates their complementary nature. Overall, the prevalence of low inter-feature correlations confirms the diversity of the features, which contributes to the discriminative ability of the identification model.

\FloatBarrier

\section{Conclusions}
This paper introduces a large-scale RF feature dataset that is constructed from $123$ same-model devices and contains $35.42$ million raw I/Q samples together with $1.85$ million extracted RF features. To establish a reproducible benchmark, we further provide a complete experimental framework that integrates data collection, feature extraction, and performance evaluation. A Random Forest–based identification model is implemented as the benchmark, which achieves an accuracy of $89.06$\% on the proposed dataset. In addition, our feature analysis reveals that frequency-related features contribute most significantly to device identification. Overall, the dataset and the framework provide a unified basis for reproducible RF fingerprinting research. They are expected to promote standardized evaluation and support future algorithmic advancements.

\section*{Declaration of competing interest}
The authors declare that they have no known competing financial interests or personal relationships that could have appeared to influence the work reported in this paper.

\section*{Acknowledgments}
This work was partly supported by JST Moonshot R\&D Grant Number JPMJMS2215, and JSPS KAKENHI Grant Number JP24K07482.

\bibliographystyle{elsarticle-num}

\bibliography{dataset}

\end{document}